\documentclass[pra,twocolumn]{revtex4}

\usepackage{enumerate}
\usepackage{amsfonts,amssymb,amsmath}
\usepackage[]{graphics,graphicx,epsfig}
\usepackage{amsthm}
\usepackage{float}

\usepackage{graphicx}
\usepackage{xcolor}
\usepackage{qcircuit}
\usepackage{tcolorbox}

\usepackage[normalem]{ulem}
\renewcommand{\underline}[1]{\uline{#1}}

\newcommand{\mk}[1]{\color{teal}{#1\,}\color{black}}

\usepackage{tikz}
\usetikzlibrary{arrows}

\usepackage{pdfpages}
\usepackage{epstopdf}
\DeclareGraphicsExtensions{.png,.pdf}

\usepackage{graphicx}
\usepackage{dcolumn}
\usepackage{natbib}
\usepackage{color}
\usepackage{multirow}
\usepackage{ulem}
\newcommand{\ket}[1]{|{#1}\rangle}

\usepackage{array}
\newcolumntype{P}[1]{>{\centering\arraybackslash}p{#1}}

\begin{document}


\title{Unitary Realizations of Synchronizing Automata in Quantum Systems}


\author{J\c{e}drzej Stempin}
\affiliation{Institute of Spintronics and Quantum Information, Faculty of Physics and Astronomy, Adam Mickiewicz University, 61-614 Pozna\'n, Poland}

\author{Jan W{\'o}jcik}
\affiliation{Institute of Spintronics and Quantum Information, Faculty of Physics and Astronomy, Adam Mickiewicz University, 61-614 Pozna\'n, Poland}

\author{Gabriela Banaszak}
\affiliation{Institute of Spintronics and Quantum Information, Faculty of Physics and Astronomy, Adam Mickiewicz University, 61-614 Pozna\'n, Poland}

\author{Andrzej~Grudka}
\affiliation{Institute of Spintronics and Quantum Information, Faculty of Physics and Astronomy, Adam Mickiewicz University, 61-614 Pozna\'n, Poland} 

\author{Marcin Karczewski}
\affiliation{Institute of Spintronics and Quantum Information, Faculty of Physics and Astronomy, Adam Mickiewicz University, 61-614 Pozna\'n, Poland}

\author{Pawe{\l} Kurzy{\'n}ski}
\email{pawel.kurzynski@amu.edu.pl}
\affiliation{Institute of Spintronics and Quantum Information, Faculty of Physics and Astronomy, Adam Mickiewicz University, 61-614 Pozna\'n, Poland}

\author{Antoni W{\'o}jcik}
\affiliation{Institute of Spintronics and Quantum Information, Faculty of Physics and Astronomy, Adam Mickiewicz University, 61-614 Pozna\'n, Poland}

\date{\today}


\begin{abstract}
We introduce a quantum analogue of a classical synchronizing automaton. In classical case the state of a system evolves according to a set of rules forming an alphabet, and sequences of these rules, called words, govern its evolution. Certain special words, known as synchronizing words, drive the automaton into a predetermined state regardless of its initial configuration. Although such an apparently irreversible process seems incompatible with the unitarity of quantum mechanics, we present a resetting protocol based on quantum synchronizing words by incorporating auxiliary qubits whose states encode the rules of the automaton’s alphabet. These qubits interact with the quantum automaton, whose state is encoded in a qudit, via a global unitary operation. When the qubit register is initially prepared in a state corresponding to a synchronizing word, the automaton evolves into a predetermined pure state independent of its initial state, while the qubit register is transformed into a complex, often entangled, state that encodes information about the automaton’s original configuration. The resulting entanglement depends on both the rule set and the automaton’s initial state, and we show how specific entangled states can be generated within this framework.
\end{abstract}

\maketitle

\section{Introduction}
Synchronization is a broad concept that appears in various fields of science. In physics it is mostly considered to be a phenomenon in which two, or more, oscillators evolve towards the same frequency and phase. However, in computer science, particularly in automata theory, synchronization is a process that takes an automaton to a predetermined state, regardless of its initial state \cite{vcerny1964poznamka,jurgensen2008synchronization}. Despite apparent differences, both notions of synchronization have one important common feature: it is a process that contracts the allowable state space of the system.

To illustrate automata synchronization, consider a robot on a 3×3 grid that can execute only two instructions: $a$ (attempt to move forward one cell; if the square ahead is outside the grid the robot stays put) and $b$ (rotate left $90^o$). A robot state is determined by its cell $(x,y)$ (with $x=0,1,2$ left $\rightarrow$ right and $y=0,1,2$ top $\rightarrow$ bottom) and its facing (one of $N$, $E$, $S$, $W$). A {\it synchronizing word} is a finite instruction sequence that, when executed from any initial state, brings every possible robot to the same final state (or — in the weaker version we use in this example — to the same cell, possibly with different facings).

\begin{figure}[t]
     \includegraphics[width=1.0\linewidth]{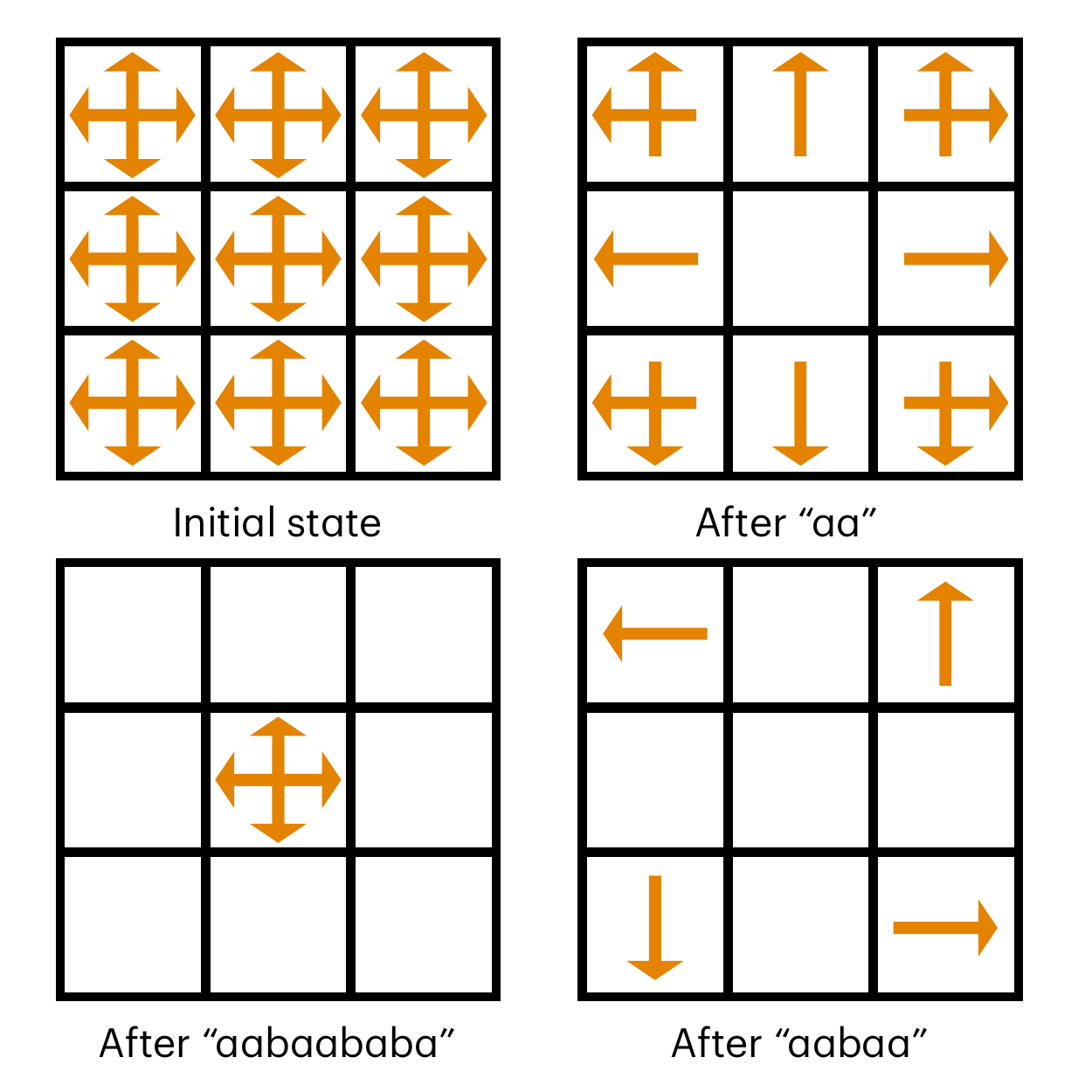}
        \caption{Schematic representation of how synchronizing word $aabaababa$ brings the robot to the center of the grid (clockwise from an arbitrary initial state -- top-left). }
    \label{robot}
\end{figure}

The nine-letter word below is a synchronizing word that sends the robot in an arbitrary initial state to the center cell (1,1) (though final facings may differ):
\begin{equation}
    aabaababa.
\end{equation}
Intuitively this sequence funnels robots inward. Because $a$ has no effect at a boundary when the robot faces outward, robots on edges that point outwards are stalled while others move; the interspersed $b$ rotations change facings so that over the whole sequence every robot will at some point face toward the interior and make the inward move. Repeating this pattern of moves-and-left-rotations eliminates the possibility of remaining on the outer ring: after the nine instructions every robot that started anywhere on the board has been moved (or held) in such a way that the only possible final cell is the center. This idea is represented in Fig. \ref{robot}.

This problem underscores the importance of a synchronization protocol -- an algorithmic sequence of $a$ and $b$ (or even more rules) that ensures convergence to a unique final state, irrespective of the initial condition. Such sequences, known as \textit{synchronizing words} \cite{ryzhikov2020synchronizing,van2022synchronizing,vcerny1964poznamka,eppstein1990reset,jurgensen2008synchronization}, are central to automata theory and have broad applications in control theory, coding, and symbolic dynamics.


\subsection*{Contribution}

In this paper, we further extend the concept of  synchronizing words to quantum systems \cite{grudka2025quantum}. A natural first approach is to translate classical synchronization schemes into the language of unitary quantum dynamics. However, a fundamental challenge arises: for synchronization to occur, there must exist operations that map multiple initial states to the same final state. This requirement conflicts with the reversibility of unitary evolution, making direct translation impossible.

More generally, irreversibility need not be a fundamental obstruction if the automaton is embedded into a suitably enlarged dynamical structure. An instructive example is provided by cellular automata: Conway’s Game of Life \cite{life} is irreversible, while the more structured Critters automaton of Toffoli and Margolus \cite{toffoli1987cellular} admits reversible evolution. Here, we propose a method for constructing quantum synchronizing automata by introducing an auxiliary system. In particular, the automaton is encoded in a qudit, while the auxiliary system is an $n$-qubit register. Each of the $n$ qubits can be in one of two orthogonal states, either $|a\rangle$ or $|b\rangle$. These states encode a specific transition rule that is applied to the qudit. We then introduce a joint qubit–qudit unitary transformation. More precisely, in the $i$-th step of the joint evolution ($i=1,\ldots,n$), the transformation acts on the $i$-th qubit and applies the corresponding transition rule to the qudit.

Quantum finite automata had been studied before, but not in the context of synchronization. Instead, they were viewed as quantum recognizers driven by a classical input word and yielding an acceptance probability \cite{qaut1,qaut2,qaut3}. Here, apart from introducing a quantum version of a synchronizing automaton, our main contribution in this work is the following proof. We show that although the transition rules $a$ and $b$ are, in general, non-unitary (since they contract the qudit state space), for certain contracting rules it is possible to construct a transformation that is unitary on the joint space. Such a transformation alters the state of the qubit register in a complex manner and, in general, produces non-trivial multi-qubit entanglement.

While our construction can be viewed as a unitary embedding of non-unitary dynamics, it differs from a generic Stinespring dilation in an essential way. In standard dilation theory, the environment is typically introduced in an abstract and non-structured manner to reproduce a given quantum channel. In contrast, here the auxiliary system has a clear computational interpretation: it encodes a word over a finite alphabet, and the global unitary is constrained to act sequentially and locally with respect to this encoding. As a result, not every dilation corresponds to a synchronizing automaton, and the existence of such a structured unitary is governed by nontrivial combinatorial constraints, captured by Theorem 1.

Our method can be used to map an arbitrary qudit state (mixed or pure) into a preassigned pure state, provided that the initial state of the qubit register is prepared in a state corresponding to a synchronizing word. This ensures reliable state preparation. Moreover, we demonstrate that, by preparing specific qudit states, our protocol can transform the qubit register into various multi-qubit entangled states. This approach naturally extends our previously proposed concept \cite{grudka2025quantum} and establishes a robust foundation for the development of new tools for quantum computation.


\section{Preliminaries -- Synchronizing Deterministic Finite Automata}


Deterministic Finite Automaton (DFA) consists of a set of possible states $Q$, a set of transition rules $\Sigma$, and a transition function $\delta: Q \times \Sigma \rightarrow Q $
that governs the system's dynamics. The set $\Sigma$ is called the alphabet and its elements are called letters. Finally, $\Sigma^{*}$ is a set of concatenated letters and its elements $w\in \Sigma^{*}$ are known as words. 

DFA can be represented as a collection of directed graphs, where a graph $G_a$ corresponds to a letter $a\in\Sigma$, its vertices correspond to states $q\in Q$, and its arcs correspond to transitions $\delta(q, a)$. 

Given an input word and an initial state, one can determine the final state. For example, given a word $aab$, where $a,b\in\Sigma$, and an initial state $q_0 \in Q$, the DFA generates $q_1 = \delta(q_0, a)$, then $q_2 = \delta(q_1, a)$, and finally $q_3 = \delta(q_2, b)$. Here we adopt a convention of reading words from left to right. 

A word $\bar{w} \in \Sigma^{*}$ that drives the automaton to the same final state, regardless of the initial state, is called a synchronizing word. If such a word exists, the corresponding automaton is said to be synchronizing. Several algorithms have been developed for identifying synchronizing words for a given DFA \cite{ryzhikov2020synchronizing,van2022synchronizing,vcerny1964poznamka,eppstein1990reset,jurgensen2008synchronization}. There is a conjecture due to {\v{C}}ern{\`y} \cite{jurgensen2008synchronization} that for a $n$-state synchronizing automaton the shortest synchronizing word is no longer than $(n-1)^2$.



\subsection*{Example 1}

\begin{figure}[h!]
  
     \includegraphics[width=0.6\linewidth]{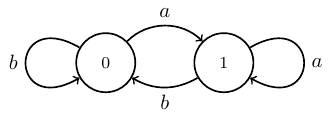}
        \caption{Example 1 -- A simple DFA can be composed of $Q = \{0,1\}$ and $\Sigma = \{a,b\}$. In this case both single letter words, $a$ and $b$, are synchronizing.}
    \label{fig:main}
\end{figure}

A simple case of a synchronizing DFA consists of two states $Q = \{0,1\}$ and two input letters $\Sigma = \{a,b\}$, as illustrated in Fig. \ref{fig:main}. Here, a single application of the letter $a$ brings the system to the state $1$, whereas a single application of the letter $b$ brings the system to the state $0$. 

\subsection*{Example 2}

\begin{figure}[h!]
         \includegraphics[width=0.85\linewidth]{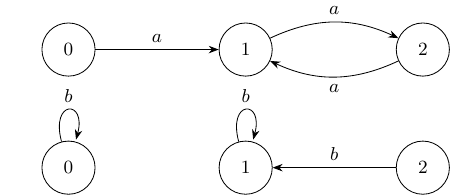}
         \vspace{3mm}
         \caption{Example 2 -- Graph corresponding to the letter $a$ (top), and to the letter $b$ (bottom). The shortest synchronizing word consists of two letters: $ab$.}
    \label{nonunit}
\end{figure}

Let us consider another simple example. We will later show that this example is of particular importance. This time a DFA consists of three states $Q = \{0,1,2\}$ and two input letters $\Sigma = \{a,b\}$, as illustrated in Fig. \ref{nonunit}. A single letter cannot bring a system to a unique state, yet the shortest synchronizing word consists of only two letters: $ab$. The letter $a$ brings the automaton to either $1$ or $2$ and the subsequent application of $b$ brings it to $1$.


\subsection*{Example 3}

\begin{figure}[h!]
         \includegraphics[width=0.65\linewidth]{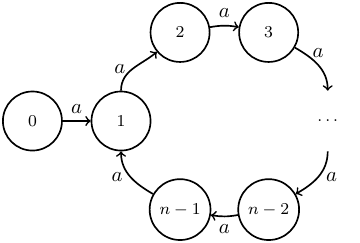}
         \includegraphics[width=0.65\linewidth]{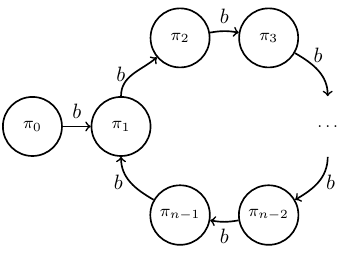}
        \caption{Example 3 -- Graph corresponding to the letter $a$ (top), and to the letter $b$ (bottom).}
    \label{n_cycles}
\end{figure}

Finally, let us consider a DFA with $n$ states. We generalize Example 1 by incorporating additional nodes into the external loop. We establish the graph corresponding to symbol $a$ as our reference, with nodes labeled in an ascending order. The graph for symbol $b$ can be derived by applying a permutation $\pi$ to the reference graph. Both graphs are depicted in Fig. \ref{n_cycles}. For the sake of simplicity, we denote permutation for which $\pi_{0} = 1$, $\pi_{1} = 0$ and $\pi_{k}=k$ for $k \geq 2$. In this case the shortest synchronizing word $\bar{w}$ is
\begin{equation}
\bar{w} \equiv (ba)^{\lfloor(n-1)/2\rfloor}  a^{(n-1)\text{mod} \hspace{0.1cm}2}.
\end{equation} 
This synchronizing word invariably directs the DFA to state $1$. Conversely, interchanging $a$ and $b$ yields a synchronizing word that terminates in state $0$. Notably, in this example, the length of the synchronizing word is precisely $n-1$.

\section{Synchronizing unitary automata}

Information processing in quantum systems is typically governed by sequences of unitary operations \cite{Nielsen_Chuang_2010}. This paradigm also applies to quantum automata (see for example work on quantum finite automata \cite{qaut1,qaut2,qaut3} and cellular automata \cite{schumacher2004reversible,perez2007local,wiesner2008quantum,grossing1988quantum,farrelly2020review,moore2000quantum}). Accordingly, we adopt this framework and make a first attempt to formulate a quantum analogue of a synchronizing DFA — a quantum DFA (QDFA) — in which state transitions are implemented by unitary operators.

First, let us start with a classical DFA and try to find its description within quantum formalism. Let $\mathcal{H}_{Q}$ be a Hilbert space corresponding to the set of automaton's states. These are represented by orthonormal vectors $\{|q\rangle\}_{q\in Q}$ that form a basis in $\mathcal{H}_{Q}$. Next, we introduce unitary operators $\{U_a\}_{a\in\Sigma}$. Therefore, words are sequences of such unitary operators. 

In particular, let us consider the DFA from Example 1 above. In this case $\mathcal{H}_{Q}$ is a two-dimensional space spanned by the basis states $\{|0\rangle,|1\rangle\}$ and the alphabet $\Sigma=\{a,b\}$ gives rise to two unitary operators $U_a$ and $U_b$. However, these operators should implement the following transformations
\begin{eqnarray}
    & U_a|0\rangle = |1\rangle,~~~~U_a|1\rangle = |1\rangle, \\
    & U_b|0\rangle = |0\rangle,~~~~U_b|1\rangle = |0\rangle,
\end{eqnarray}
which are clearly in conflict with the assumed unitarity.

The above issue will also arise in Examples~2 and~3. More generally, this problem occurs whenever the DFA transition rule is contracting. Consequently, we must devise a way to circumvent it. Here, we adopt a method that has been successfully applied in the context of quantum walks~\cite{QW1,QW2}. Specifically, we introduce an additional Hilbert space to encode information about the transition rules.

Consider a QDFA that processes a \(k\)-letter word. The state of the automaton is described as before. However, in this setting, each letter of the word is encoded in a qudit ($d=|\Sigma|$) and represented by a vector \(\{|a\rangle\}_{a\in\Sigma}\). The composite system, consisting of the automaton and the input word, is therefore represented by a vector in the product space \(\mathcal{H}_{\Sigma}^{\otimes k} \otimes \mathcal{H}_{Q}\), where $\mathcal{H}_{\Sigma}$ is the Hilbert space corresponding to a single letter. In the remainder of this work, we restrict attention to two-letter alphabets $\Sigma=\{a,b\}$; consequently, letters are encoded in qubits and \(\dim(\mathcal{H}_{\Sigma}) = 2\). The corresponding word is therefore encoded in a $k$-qubit product state, e.g., 
\begin{equation}    |a\rangle_1|b\rangle_2|a\rangle_3\ldots|a\rangle_k \equiv |aba\ldots a\rangle,
\end{equation} 
where, for brevity, we skipped the tensor product symbol.

\begin{figure}[t] \includegraphics[width=1.0\linewidth]{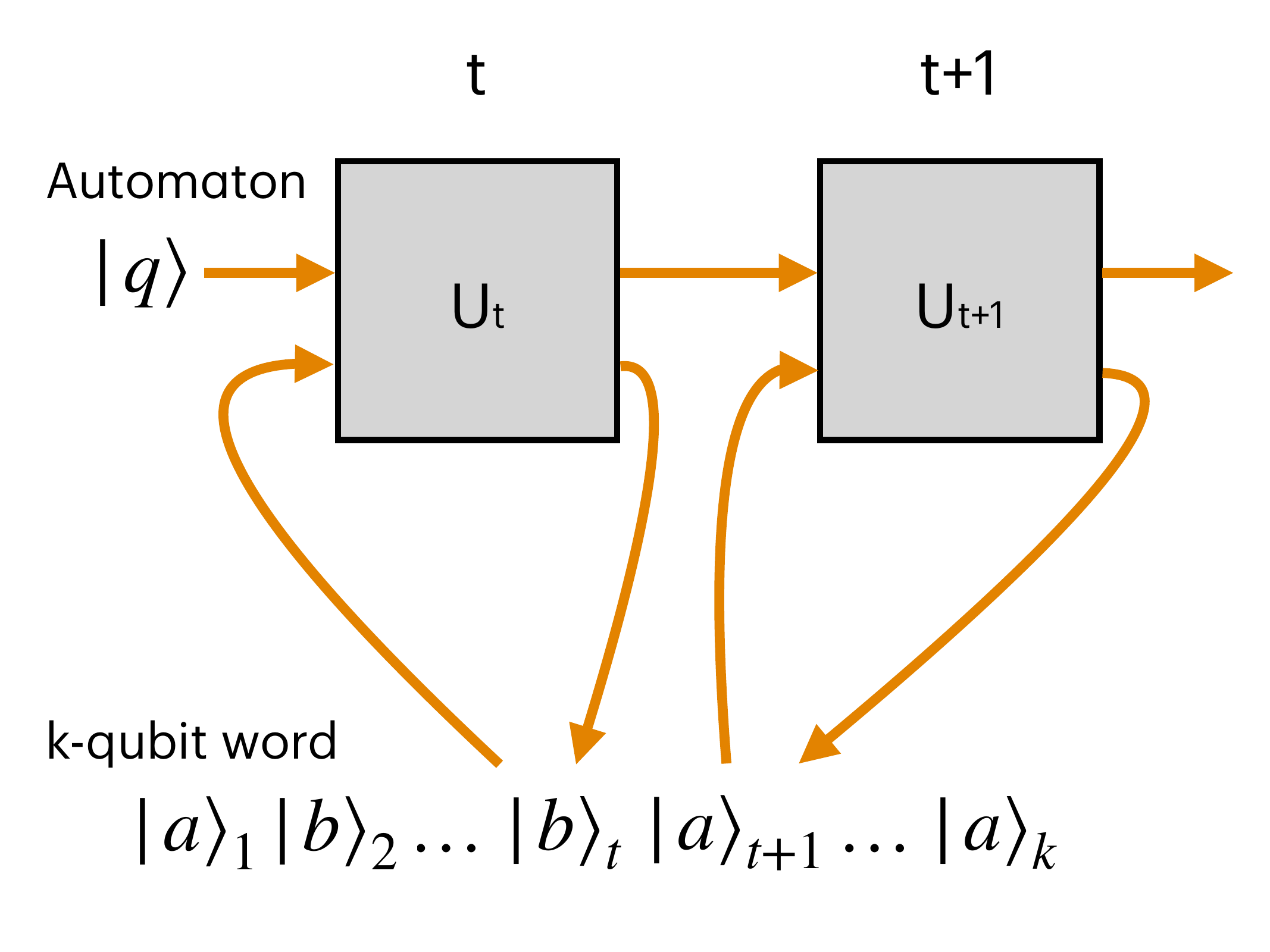}
        \vspace{-10mm}
        \caption{Schematic representation of two steps (step $t$ and $t+1$) of a unitary QDFA that processes a k-qubit word $ab\ldots ba\ldots a$. }
    \label{U}
\end{figure}

Next, let $U_j$ be a unitary operation acting on the qudit that encodes the automaton’s state and on a qubit corresponding to a $j$'th letter. More precisely, at time steps $t = 1,2,\dots,k$, the operation $U_t$ acts on the automaton and the $t$-th qubit of the $k$-qubit word. This process is schematically represented in Fig.~\ref{U}. The main problem is -- which automata admit such a unitary description? Before providing a general answer, let us reconsider the above three examples. 

\subsection*{Example 1}

In this example the states of the automaton are $\{|0\rangle,|1\rangle\}$ and we need to consider the action of $U_j$ on four composite states $\{|a\rangle_j|0\rangle,|a\rangle_j|1\rangle,|b\rangle_j|0\rangle,|b\rangle_j|1\rangle\}$. Since we want to recover the action of classical DFA, i.e., we do not wish to generate superpositions, the action of $U_j$ should be a permutation of the four composite states. This can be achieved by the following $U_j$
\begin{eqnarray}
& &U_j|a\rangle_j|0\rangle = |a\rangle_j|1\rangle, \\
& &U_j|a\rangle_j|1\rangle = |b\rangle_j|1\rangle, \\
& &U_j|b\rangle_j|0\rangle = |a\rangle_j|0\rangle, \\
& &U_j|b\rangle_j|1\rangle = |b\rangle_j|0\rangle.
\end{eqnarray}
The idea behind this operation is that the automaton’s state is updated according to the corresponding rule, and the letter’s state, i.e., the state of the corresponding qubit, is swapped if the automaton’s state does not change.

As already mentioned earlier, the above automaton is synchronizing, and the synchronization property carries over to the quantum regime. In particular, consider an initial state of the automaton to be an arbitrary superposition $\alpha |0\rangle + \beta|1\rangle$. The following action of $U_j$
\begin{equation}
    U_j|a\rangle_j(\alpha |0\rangle + \beta|1\rangle)=(\alpha |a\rangle_j + \beta|b\rangle_j)|1\rangle
\end{equation}
brings the automaton to the state $|1\rangle$, whereas the superposition is swapped to the letter's state. The above synchronization to state $|1\rangle$ will also work if the automaton were initially in an arbitrary mixed state.


\subsection*{Example 2}

This time the states of the automaton are $\{|0\rangle,|1\rangle, |2\rangle\}$ and we need to study how $U_j$ acts on $\{|a\rangle_j|0\rangle,|a\rangle_j|1\rangle,|a\rangle_j|2\rangle,|b\rangle_j|0\rangle,|b\rangle_j|1\rangle,|b\rangle_j|2\rangle\}.$ In general we should have
\begin{eqnarray}
& &U_j|a\rangle_j|0\rangle = |\cdot\rangle_j|1\rangle, \\
& &U_j|a\rangle_j|1\rangle = |\cdot\rangle_j|2\rangle, \\
& &U_j|a\rangle_j|2\rangle = |\cdot\rangle_j|1\rangle, \\
& &U_j|b\rangle_j|0\rangle = |\cdot\rangle_j|0\rangle, \\
& &U_j|b\rangle_j|1\rangle = |\cdot\rangle_j|1\rangle, \\ 
& &U_j|b\rangle_j|2\rangle = |\cdot\rangle_j|1\rangle,
\end{eqnarray}
where $\cdot$ denotes some letter state. However, as already mentioned, $U_j$ should permute basis states, but in the above $U_j$ generates four states of the form $|\cdot\rangle_j|1\rangle$. These four states must be mutually orthogonal, yet we have only two possible options -- either $|a\rangle_j|1\rangle$ or $|b\rangle_j|1\rangle$. Therefore, in our setting there is no $U_j$ corresponding to the DFA from Example 2. 


\subsection*{Example 3}

Finally, we show that DFA from Example 3 can be unitarized. Consider $U_j$ that acts on $$\{|a\rangle_j|0\rangle,\ldots,|a\rangle_j|n-1\rangle,|b\rangle_j|0\rangle,\ldots,|b\rangle_j|n-1\rangle\}$$ and generates the following permutation:
\begin{equation}
    \begin{array}{ccc}
      \ket{a}_{j}\ket{0} & \leftarrow  & \ket{b}_{j}\ket{\pi_{n-1}}\\
      \downarrow &  & \uparrow \\
       \ket{a}_{j}\ket{1} &  & \ldots\\
      \downarrow & & \uparrow\\
       \ket{a}_{j}\ket{2} & & \ket{b}_{j}\ket{\pi_{2}} \\
      \downarrow & & \uparrow\\
      \ldots & & \ket{b}_{j}\ket{0} \\
      \downarrow & & \uparrow\\
      \ket{a}_{j}\ket{n-1} & \rightarrow & \ket{b}_{j}\ket{1} \\
    \end{array}
    \label{cycle}
\end{equation}
This permutation implements the correct DFA transition rules. 

Let us focus on a specific case, discussed in previous section, where $\pi_{0} = 1$, $\pi_{1} = 0$ and $\pi_{k}=k$ for $k \geq 2$. More precisely, let us consider $n=4$ and the automaton initialized in an arbitrary superposition. Next, let us prepare a three-qubit register in the state $\ket{a}_{1}\ket{b}_{2}\ket{a}_{3}$ and apply the transformation:
\begin{equation}
\begin{array}{c}
   U_{3}U_{2}U_{1} \ket{a}_{1}\ket{b}_{2}\ket{a}_{3} \otimes (\alpha \ket{0} + \beta \ket{1} + \gamma \ket{2} + \delta \ket{3})=      \\
   \\
      =(\alpha \ket{aba} + \beta\ket{abb} + \gamma\ket{aaa} + \delta\ket{bba})\otimes \ket{1}.
\end{array}
\end{equation}
The final state of the automaton is $|1\rangle$, no matter what the initial superposition was. Therefore, the word encoded in the initial qubit register is a synchronizing word. Note, that after the transformation the information about the automaton's initial state is encoded in the complex multi-qubit state, which in general can be genuinely multipartite entangled. We will come back to this problem in a moment.   


\subsection*{Which automata can be unitarized?}

Now we focus on a general problem -- which DFAs can be unitarized within our model? Our construction is a Stinespring dilation constrained in two ways: the environment is an $n$-qubit register encoding words in $\Sigma^n$, and the global unitary must factor as $U_k \cdots U_1$ with $U_j$ acting only on the automaton and the $j$-th qubit. Crucially, the register storing the input word doubles as the dilation environment, so no additional ancillary space is required. The following theorem reduces the existence of such a dilation to a combinatorial balance condition on the transition graphs.

 \begin{tcolorbox}[
        colback=gray!10!white, 
        colframe=gray!50!white, 
        sharp corners, 
        boxsep=0.5mm, 
        arc=0mm, 
        boxrule=0.2mm 
    ]
    \noindent {\bf Theorem 1.} There exists a unitary transformation $U_j$ if and only if for each state $q\in Q$ the total number of both, outgoing and incoming arcs, summed over all graphs $G_a$ ($a\in \Sigma$), equals to $|\Sigma|$.
    \end{tcolorbox}

\begin{proof}
We look for a unitary transformation $U_j$ on basis states $\{|a\rangle_j|q\rangle\}_{a\in\Sigma,q\in Q}$ (where in general $|\Sigma|\geq 2$ and $|Q|=n$) that implements the rules of the corresponding DFA. Such a transformation must act as a permutation of this basis (up to a local unitary transformation of the letter basis, which we can skip without a loss of generality). In a permutation, each basis state has exactly one pre-image. Let us consider all basis states whose automaton's component is $q$  
\begin{equation}
\ket{a}\ket{q},\ \ket{b}\ket{q},\ \ldots,\ \ket{z}\ket{q}.
\end{equation}
Each of these states must have exactly one pre-image under the transformation. Therefore, in the graph representation, for each vertex $q$, the total number of incoming arcs, summed over all graphs $G_{a}$ corresponding to different letters $a\in\Sigma$, is $|\Sigma|$. Likewise, the total number of outgoing arcs from each vertex, also summed over all letters, must be equal to $|\Sigma|$, and hence equal to the total number of incoming arcs.

On the other hand, if for each vertex $q$ the sum  of all outgoing (incoming) arcs over all graphs is equal to $|\Sigma|$, then one can associate a unique distinct image (pre-image) $\ket{a}\ket{q}$ to each arc. Therefore, the corresponding transformation is a permutation and hence is unitary. 
\end{proof}

Although the condition in Theorem 1 has a simple combinatorial form, its significance lies in providing a complete characterization of when classical, generally non-invertible transition rules can be embedded into a reversible quantum evolution under the structural constraints of our model. In particular, it identifies a precise boundary between automata that admit a unitary realization and those that fundamentally do not, thereby linking synchronization theory with reversibility constraints in quantum dynamics.

This boundary is sharpened by a clear graph-theoretic interpretation. Consider the labeled directed multigraph $G = \bigcup_{a\in\Sigma} G_a$ obtained by superimposing all transition graphs on the common vertex set $Q$, with each arc labelled by the corresponding letter. Since every state has exactly one outgoing arc per letter, $G$ has uniform out-degree $|\Sigma|$ at every vertex, and Theorem~1 asserts that a unitary realization exists precisely when the in-degrees match, i.e., when $G$ is \emph{balanced}. Balance and connectivity are the two conditions for the existence of an Eulerian circuit, so whenever $G$ is also connected, such an Eulerian circuit provides an explicit construction of $U_j$.

Let us revisit Examples 1-3 for the last time and examine them in the context of the above theorem. Note that for Examples 1 and 3 the sum of incoming and outgoing arcs for each vertex is two, hence a proper unitarization exists. On the other hand, for Example 2 the sum of incoming (outgoing) arcs to (from) vertices 0 and 2 is one (two), whereas the sum of incoming (outgoing) arcs to (from) vertex 1 is three (two). Therefore, there is no unitarization in this case.


\subsection*{How many DFAs can be unitarized?}

Let us consider $n$-state DFAs over the two-letter alphabet $\Sigma = \{a,b\}$. How many such DFAs are there, and what fraction of them can be unitarized using our method?

First, we note that, just as in Fig.~\ref{fig:main}, each such automaton can be represented as a graph in which each vertex has exactly two outgoing arcs. Moreover, each arc is labeled either by the letter $a$ or $b$, so all arcs are distinguishable, and there are $2n$ such arcs in total. Each arc can point to any of the $n$ possible vertices, hence there are
\begin{equation}
    N_{DFA}(n)=n^{2n},
\end{equation}
possible DFAs.

Next, according to Theorem~1, each unitarizable DFA over a two-letter alphabet corresponds to a subset of the above graphs in which each vertex has exactly two outgoing arcs and two incoming arcs. Such a DFA is equivalent to a distribution of the $2n$ distinguishable arcs (labeled $q_a$ and $q_b$ for each vertex $q$) into $n$ distinguishable target vertices, with exactly two arcs per target. The number of such distributions is the multinomial coefficient
\begin{equation}
    N_{\text{QDFA}}(n) = \binom{2n}{2, 2, \ldots, 2} = \frac{(2n)!}{2^{n}}.
    \end{equation}
    Thus, the fraction of DFAs that can be unitarized is
\begin{equation}
    N_{QDFA}(n)=\frac{(2n)!}{2^{n}}.
\end{equation}

Finally, the fraction of DFAs that can be unitarized is
\begin{equation}
    f_{QDFA}(n)=\frac{N_{QDFA}(n)}{N_{DFA}(n)}=\frac{(2n)!}{2^{n} n^{2n}}.
\end{equation}
This expression can be approximated using Stirling's formula:
\begin{equation}
    f_{QDFA}(n)\approx \sqrt{4\pi n}\left(\frac{2}{e^2}\right)^n,
\end{equation}
where $2/e^2 \approx 0.27$. Hence, for large $n$, the fraction of unitarizable DFAs decreases exponentially.


\section{Three types of QDFA's dynamics}

\subsection*{Assumptions}

\begin{figure}[h!]
    \centering
    \includegraphics[width=1\linewidth]{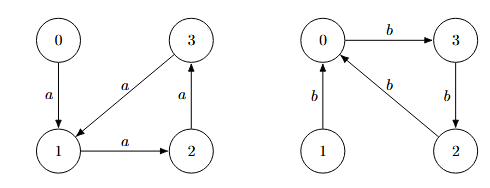}
    \caption{Transition graphs corresponding to the action of letters $a$ and $b$.}
    \label{QDFA1}
\end{figure}

\begin{figure}[h!]
    \centering
    \includegraphics[width=0.8\linewidth]{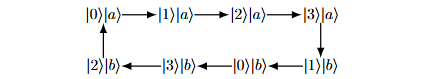}
    \caption{Diagram representing unitary QDFA state transitions corresponding to transition graphs from Fig. \ref{QDFA1}.}
    \label{QDFA2}
\end{figure}

As mentioned above, we focus on alphabets consisting of just two letters $\Sigma=\{a,b\}$, which are encoded in qubits. These letters are concatenated into words of finite length $k$, hence we consider $k$-qubit registers and the entire evolution consists of exactly $k$ steps. In addition, we assume that the initial state of the system is a product state of the automaton and the qubit register, and that the qubit register itself is a product of basis states; that is, it encodes a classical word such as $\ket{a}_1\ket{b}_2\ldots\ket{a}_k$. Given the above assumptions, we identify three different QDFA behaviors. To illustrate them, throughout this section we use an exemplary QDFA with $n=4$ states, whose evolution is presented in Figs. \ref{QDFA1} and \ref{QDFA2}. It is a particular case of the automaton in Example 3. 

\subsection*{Behavior 1: Emulation of classical DFA}

The first behavior is a simple consequence of our QDFA model. Let us recall that the unitarization introduced in the previous section generates permutations of basis states. As a consequence, the action on a basis state does not generate any superposition. Therefore, if the system is initialized in a basis state, which can be interpreted as a classical state, it will remain a classical state throughout the entire evolution. Such an evolution can be considered as a quantum emulation of the corresponding classical DFA.


\subsection*{Behavior 2: Automaton-register entanglement}

Now, let us consider a quantum situation in which the initial state is a superposition of basis states. For example, it can be a product state of the automaton in a superposition of basis states and the qubit register encoding a classical word. This situation was already considered in the previous section; however, we focused on a particular case -- the qubit register was encoding a synchronizing word. If the encoded word is not synchronizing, one may end up in an entangled state of the automaton and the register. This situation is represented below using the exemplary automaton from Figs. \ref{QDFA1} and \ref{QDFA2}.
\begin{equation}
\begin{array}{c}
  U_{3}U_{2}U_{1} \ket{a}_{1}\ket{a}_{2}\ket{b}_{3} \otimes (\alpha \ket{0} + \beta \ket{1} + \gamma \ket{2} + \delta \ket{3})=      \\
   \\
      =(\alpha \ket{aaa}+\gamma\ket{abb} + \delta\ket{baa})\ket{0} + \beta\ket{aab}\ket{2}.
\end{array}
\end{equation}
Generation of such entanglement can be potentially useful for various quantum information processing tasks. In general, this type of behavior is rich and complex and deserves more detailed analysis. However, in this work we focus on the last type of behavior that leads to particularly interesting outcomes.  


\subsection*{Behavior 3: Automaton-register decoupling}

The last behavior corresponds to the situation in which the register-automaton system ends up in a product state, and one may independently study the final states of the register and the automaton. The synchronization that has already been discussed is a prime example of such behavior. As already mentioned, due to quantum synchronization, the initial product state of the register that encodes the synchronizing word can evolve into a complex entangled state. 

At least two scenarios can be considered. In a standard synchronization scenario the control over initial automaton's state is assumed to be either lost, or limited, and a proper choice of the qubit register's initial state allows the automaton to evolve into a known pre-determined state at the cost of swapping the initial randomness of the automaton onto qubits. On the other hand, one may have full control over both initial states and the goal would be to exploit the system's dynamics in order to achieve some desired state. In particular, since the register is itself a composite system, one may attempt to generate various entangled multi-qubit states, such as GHZ (Greenberger-Horne-Zeilinger) \cite{Greenberger1989}, W \cite{PhysRevA.62.062314} or AME (Absolutely-Maximally-Entangled) \cite{Helwig_2012} states, determined by the preparation of the automaton’s initial state, structure of the evolution and by the sequence of letters in the synchronizing word.

\subsection*{Entropic perspective}

For pure inputs, the three behaviors admit a simple entropic interpretation. Let $|\Psi_{\text{in}}\rangle = |w\rangle_R\otimes|\psi\rangle_Q$, $|\Psi_{\text{out}}\rangle = U_k\cdots U_1|\Psi_{\text{in}}\rangle$ and denote by $S_R$ and $S_Q$ the von Neumann entropies of the register and automaton reduced states. By purity, $S_R = S_Q$ at all times. Behaviors 1 and 3 both have initial and final $S_R = 0$ and are distinguished by whether the final state is a basis vector (Behavior 1) or a non-trivial product $|\phi\rangle_R\otimes|q_\star\rangle_Q$ (Behavior 3), in which case the initial coherence on $Q$ has been transferred into the structure of $|\phi\rangle_R$. Behavior 2 has final $S_R > 0$: the automaton and the register end entangled, with mutual information $I(Q{:}R) = 2S_R$ measuring the correlations.

The picture becomes sharper for mixed inputs, where synchronization acts as a coherent entropy pump: if $\rho_Q$ is arbitrary and the word is synchronizing, unitary invariance of entropy together with the purity of the final automaton state ensures that
\begin{equation}
S(\rho_R^{\text{out}}) = S(\rho_Q).
\end{equation}

Thus, the automaton's initial entropy is fully transferred into the register.


\section{Generation of multi-qubit entanglement}

Now we switch the scope of our interest from the final state of the automaton to the final state of the register. We feed the automaton with a special word and analyze how its particular initial state gives rise to a multi-qubit entanglement in the register. 

The constructions presented in this section are intended as explicit demonstrations of the expressive power of the framework rather than as an exhaustive classification. Our goal is to illustrate that synchronizing automata can be systematically used as generators of multipartite entanglement, with the structure of the automaton determining the class of states that can be produced.


\subsection*{Generation of bipartite entanglement and tripartite GHZ entanglement}

Let us reconsider the automaton from Figs. \ref{QDFA1} and \ref{QDFA2}. Consider the following evolution 
\begin{equation}
\begin{array}{c}
U_4 U_3 U_2 U_1 \ket{a}_1\ket{b}_2\ket{b}_3\ket{a}_4 \otimes (\beta \ket{2} + \delta \ket{0}) = \\
\\
\ket{a}_1\ket{b}_2 \otimes (\beta\ket{a}_3\ket{a}_4 + \delta \ket{b}_3\ket{b}_4) \otimes \ket{1}.
\end{array}
\end{equation}
The final state of the first and the second qubit in the register is entangled, whereas the remaining parts of the system are decoupled. 

Next, consider the same automaton, fed with the same synchronizing word but starting in a different state 
\begin{equation}
\begin{array}{c}
U_4 U_3 U_2 U_1 \ket{a}_1\ket{b}_2\ket{b}_3\ket{a}_4 \otimes (\gamma \ket{2}+ \delta \ket{3}) = \\
\\
\ket{b}_1 \otimes (\gamma \ket{a}_2\ket{a}_3\ket{a}_4 + \delta \ket{b}_2\ket{b}_3\ket{b}_4) \otimes \ket{1}.
\end{array}
\end{equation}
This time, the evolution results in a tripartite GHZ state of the second, third, and fourth qubit. 

We observed an interesting feature -- one can use the same automaton to generate different types of entanglement. This can be controlled by making a proper choice of the initial state. We will discuss this in the last subsection. 


\subsection*{Generation of tripartite W entanglement}

\begin{figure}[h!]
    \centering
    \includegraphics[width=1\linewidth]{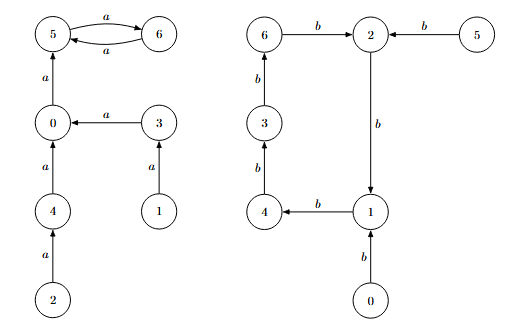}
    \caption{Transition graphs corresponding to the action of letters $a$ and $b$ on the automaton's states used for W state generation.}
    \label{fig8}
\end{figure}

\begin{figure}[h!]
    \centering
    \includegraphics[width=1\linewidth]{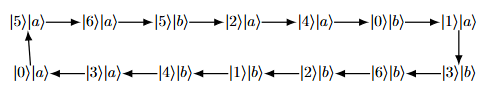}
    \caption{Diagram representing unitary QDFA state transitions corresponding to transition graphs from Fig. \ref{fig8}.}
    \label{fig9}
\end{figure}

To generate a 3-qubit W state, we need to introduce a more complex automaton with a new set of transition rules, shown in Figs. \ref{fig8} and \ref{fig9}.
The following three-step evolution leads to a generation of a tripartite W state
\begin{equation}
\begin{array}{c}
U_3 U_2 U_1 \ket{a}_1\ket{a}_2\ket{a}_3 \otimes (\alpha \ket{0} + \beta \ket{1} + \gamma \ket{2}) = \\
\\
(\alpha \ket{a}_1\ket{a}_2\ket{b}_3+\beta \ket{a}_1\ket{b}_2\ket{a}_3+\gamma \ket{b}_1\ket{a}_2\ket{a}_3) \otimes \ket{5}.
\end{array}
\end{equation}


\subsection*{Generation of AME state}

\begin{figure}[h!]
    \centering
    \includegraphics[width=0.8\textwidth,trim=0 0 100 0,clip]{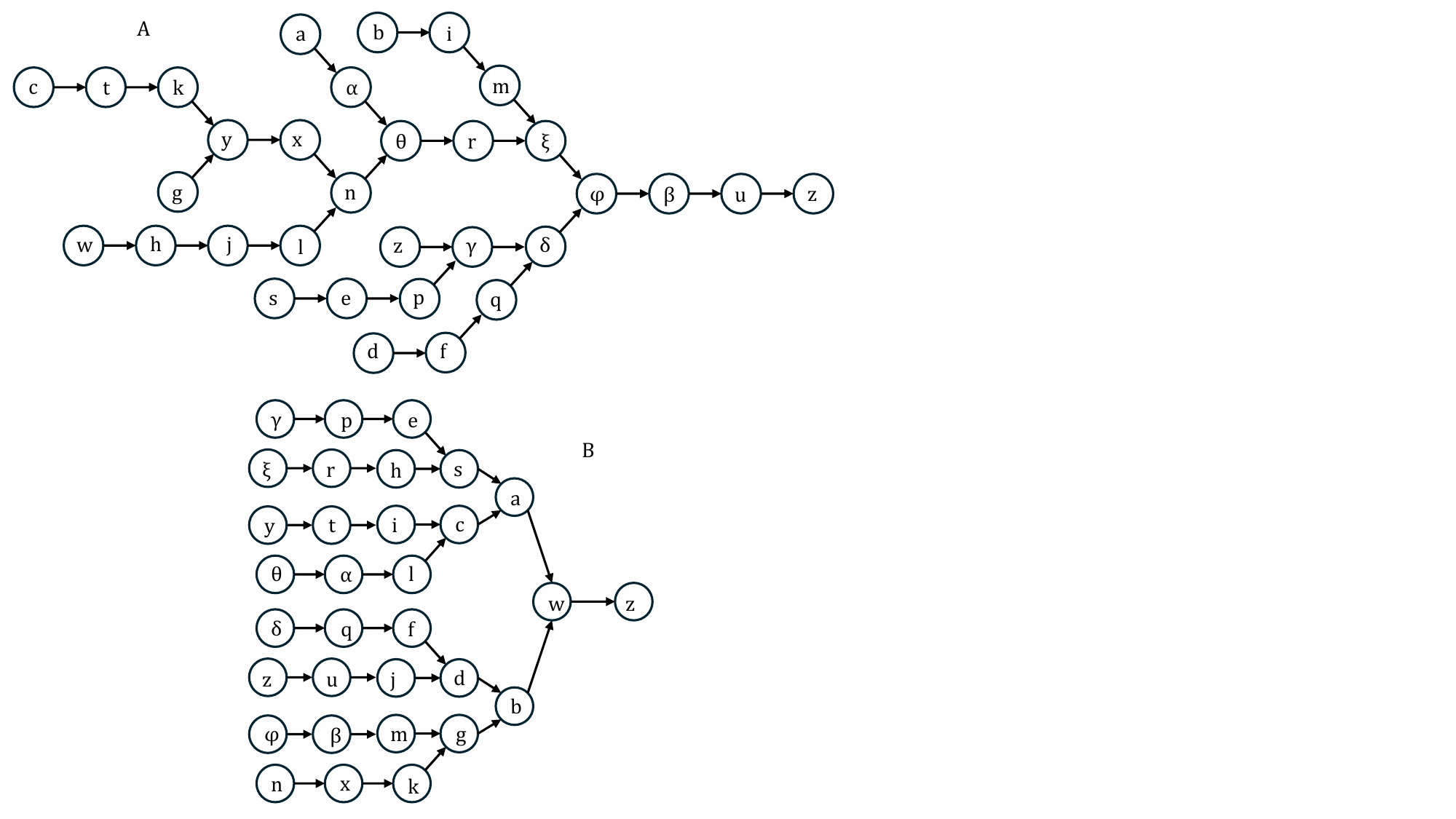}
    \caption{Transition graphs corresponding to the action of letters $A$ and $B$ on the automaton's states used for $5$-qubits AME state generation. Note that for clarity the state $z$ in both graphs appears twice - once as a vertex without outgoing arcs and once as a vertex without incoming arcs. Therefore, $z$ has one incoming and one outgoing arc.}
    \label{fig10}
\end{figure}

\begin{figure}[h!]
    \centering
    \includegraphics[width=0.6\textwidth,trim=10 140 450 40,clip]{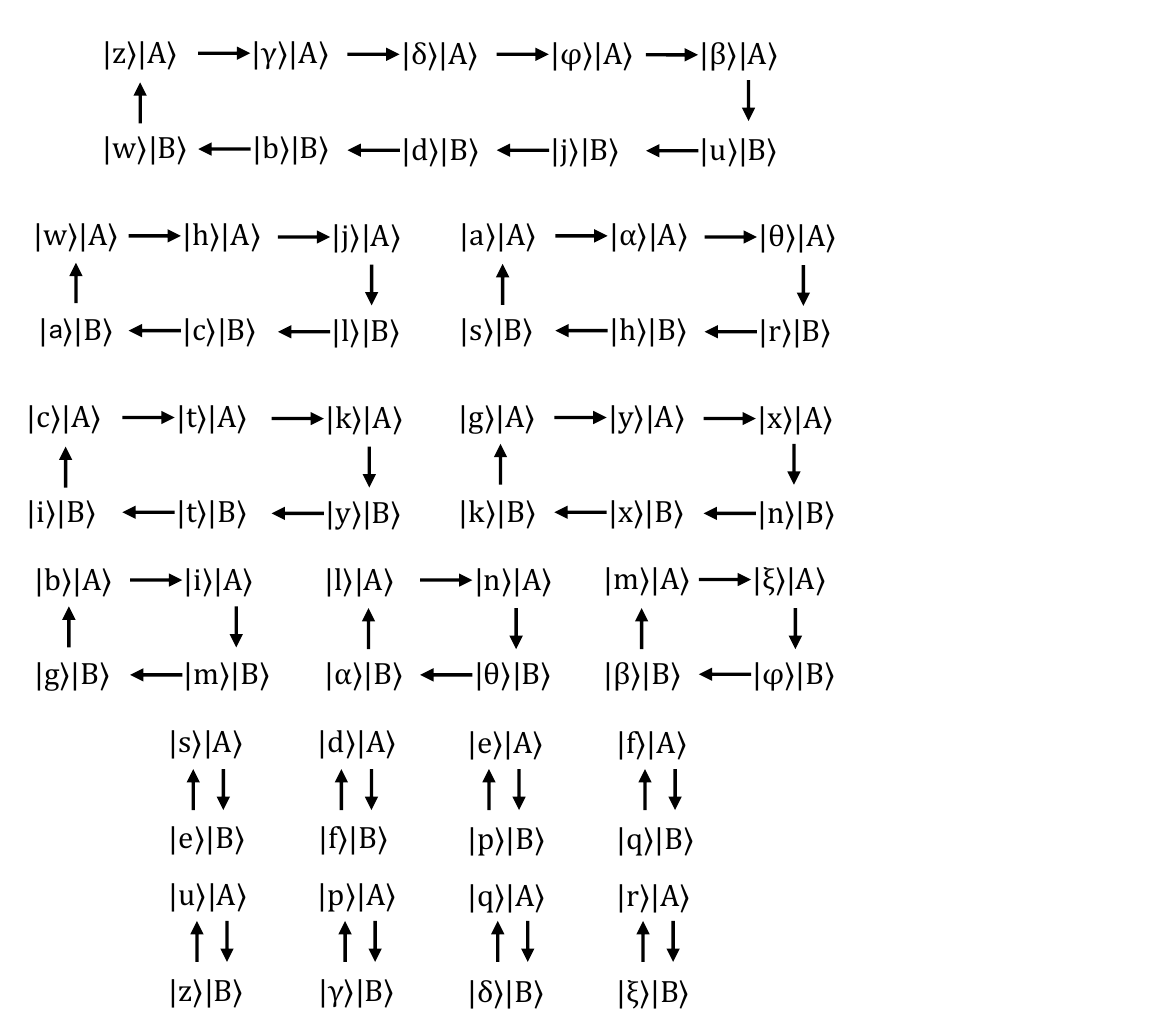}
    \caption{Diagram representing unitary QDFA state transitions corresponding to transition graphs from Fig. \ref{fig10}.}
    \label{fig11}
\end{figure}

Finally, let us introduce an automaton capable of generating a five-qubit AME state
\begin{align}
&\frac{1}{2\sqrt{2}}\big(
\ket{AAAAA} + \ket{AAABB} + \ket{ABBAA} + \ket{BBABA} \notag \\
&- \ket{ABBBB} + \ket{BBAAB} + \ket{BABBA} - \ket{BABAB}
\big).
\end{align}
A pure $n$-qubit state is AME if 
every reduction to $\lfloor n/2 \rfloor$ qubits is maximally mixed. We managed to find a 31-state unitary QDFA (see Figs.~\ref{fig10} and \ref{fig11}), whose states are labeled using letters from the Latin and Greek alphabets $Q=\{a,b,\ldots,\varphi\}$ ($|Q|=31$). To avoid confusion, this time the alphabet consists of capital letters $\Sigma=\{A,B\}$.

The five-step evolution of this QDFA transforms the following initial state
\begin{eqnarray}
& &
|BBBBB\rangle \otimes \label{ame} \\
& &
\frac{1}{2\sqrt{2}}
\Big(
|\gamma\rangle
+ |\delta\rangle
+ |\xi\rangle
+ |y\rangle
- |z\rangle
+ |n\rangle
+ |\theta\rangle
- |\varphi\rangle
\Big), \nonumber
\end{eqnarray}
into
\begin{align}
\big(
&\ket{AAAAA} + \ket{AAABB} + \ket{ABBAA} + \ket{BBABA} \notag \\
&- \ket{ABBBB} + \ket{BBAAB} + \ket{BABBA} - \ket{BABAB}
\big) \notag \\
&\otimes \frac{1}{2\sqrt{2}}|w\rangle.
\end{align}
For clarity, we have omitted the qubit labels from the formulas above. The resulting automaton is remarkably complex and was derived through a process of educated guess. The core intuition behind this approach was that the AME state (\ref{ame}) consists of a superposition of eight terms; consequently, its generation requires the synchronization of eight distinct automaton states. This explains the tree-like structure of graph $B$ in Fig. \ref{fig10}, where eight branches converge at the root $w$ , resulting in the synchronizing word $BBBBB$.


\subsection*{Universal aspects of complex automata}

A full classification of the entanglement classes accessible within this framework, as well as the identification of minimal automata generating a given state, remains an open problem and constitutes an interesting direction for future work. Nevertheless, in the above examples, we observed that generation of different entangled states requires QDFAs of different complexities. For example, tripartite GHZ states could be generated with a four-state QDFA, whereas generation of tripartite W states requires more complex automata. Here we found one with 7 states, although we do not prove that this is optimal -- we simply couldn't find a simpler QDFA that achieves this goal. 

Interestingly, the W-state QDFA (Figs. \ref{fig8} and \ref{fig9}) is capable of producing a tripartite GHZ state. Consider the following evolution
\begin{equation}
\begin{array}{c}
U_3 U_2 U_1 \ket{a}_1\ket{a}_2\ket{a}_3 \otimes (\alpha \ket{2} + \beta \ket{6}) = \\
\\
(\alpha\ket{a}_1\ket{b}_2\ket{a}_3 + \beta \ket{b}_1\ket{a}_2\ket{b}_3) \otimes \ket{5}.
\end{array}
\end{equation}
Applying a local Pauli-$X$ transformation on the middle qubit maps this state to
\begin{equation}
\alpha\ket{aaa}+\beta\ket{bbb},
\end{equation}
which is the standard form of the three–qubit GHZ state. Moreover, the automaton of Figs. \ref{fig8} and \ref{fig9} generates a GHZ state more efficiently than the automaton of Figs. \ref{QDFA1} and \ref{QDFA2}, because it does so in three, not four, steps and does not require auxiliary qubits. In addition, both the GHZ and the W state, were generated using the same word -- $aaa$. 

The above naturally leads to the following conjecture -- the more complex the QDFA is, the more universal it gets, in the sense that the more classes of entangled states it can generate. Therefore, one can hope that the AME state automaton (Figs. \ref{fig10} and \ref{fig11}) can be used to generate other classes of entangled states. Indeed, below we show that it can be used to generate GHZ \mk{and W} states.

A more formal version of the problem related to our conjecture is the following: given an $n$-state QDFA and a synchronizing word of length $m$, what is the maximum entanglement depth -- that is, the largest number of qubits that are genuinely multipartite entangled -- achievable on an $m$-qubit register, and which entanglement classes are accessible, viewed as independent functions of $n$ and $m$? Note that the register length need not equal the length of the shortest synchronizing word, since one may use longer words tailored to a specific entanglement target. Thus, $n$ and $m$ are genuinely independent parameters.

We emphasize that our constructions are not optimized with respect to the number of automaton states. In particular, the 31-state automaton used for AME state generation should be viewed as an existence proof rather than an optimal solution. 

The AME state automaton was designed to generate the five–qubit AME state, which is a superposition of eight computational basis strings. Because several of these strings differ only on a subset of qubits while the remaining qubits are identical, appropriate superpositions of the corresponding initial automaton states allow GHZ states to appear on subsets of qubits.

For example, the strings
\begin{equation}
\ket{AAAAA}, \qquad \ket{ABBBB}    
\end{equation}
agree on the first qubit and differ on qubits $2$--$5$. Consequently, if instead of state (\ref{ame}) the system was initialized in the state
\begin{equation}
|BBBBB\rangle \otimes \frac{1}{\sqrt{2}}
\Big(
|\gamma\rangle - |z\rangle \Big), \nonumber
\end{equation}
it would evolve into
\begin{equation}
\frac{1}{\sqrt{2}} \big(\ket{AAAAA} - \ket{ABBBB} \big) \otimes |w\rangle,
\end{equation}
where qubits $2$--$5$ are in the GHZ state. Similarly, other pairs of strings in the AME support differ on exactly three qubits while the remaining qubits act as spectators, giving rise to three–qubit GHZ states.

A W-state protocol within the same automaton and the same unitarization is also possible. Consider the transformation
\begin{align}
&U_4 U_3 U_2 U_1 \ket{BABB} \otimes \frac{1}{\sqrt{3}}\big(\ket{a} + \ket{e} + \ket{g}\big) \nonumber \\
&=\ket{A}_{1} \frac{\ket{A}_{2}\ket{B}_{3}\ket{A}_{4} + \ket{B}_{2}\ket{A}_{3}\ket{A}_{4} + \ket{A}_{2}\ket{A}_{3}\ket{B}_{4}}{\sqrt{3}} \otimes \ket{a}.
\end{align}
The first qubit factors out as a spectator in state $\ket{A}$, and qubits $2$--$4$ realize the standard tripartite W state. Notably, this construction uses a different initial superposition, a different word, and a different synchronization target than the AME protocol of Eq.~(\ref{ame}), yet is compatible with the same choice of $U_j$. This means that the unitarization fixed by the AME requirements leaves enough freedom to also support W state generation. This is a concrete instance of our broader conjecture: a QDFA designed for one entanglement class may, through its complex combinatorial structure, support others as well.


\section{Conclusions}

In this work, we introduced a quantum analogue of synchronizing deterministic finite automata by embedding classical transition rules into an extended Hilbert space and implementing their action through global unitary dynamics. By encoding input words into an auxiliary qubit register, we showed that processes that are inherently contracting in the classical setting can be realized within a fully unitary quantum framework, with the apparent loss of information being transferred to the auxiliary system.

Our main theoretical result establishes a necessary and sufficient condition for the unitarizability of a classical automaton. Specifically, we showed that a unitary realization exists if and only if, for each state, the total number of incoming and outgoing transitions—summed over all letters of the alphabet—is balanced. This provides a clear structural criterion linking graph-theoretic properties of automata with the constraints of quantum evolution. Moreover, we calculated the fraction of classical DFAs that can be unitaraized.

Within this framework, we demonstrated that quantum synchronizing words enable reliable state preparation: an arbitrary (pure or mixed) initial state of the automaton can be driven to a predetermined pure state, while the information about the initial configuration is coherently transferred to the auxiliary register. This mechanism can be viewed as a unitary analogue of state reset, where irreversibility is effectively displaced rather than eliminated.

Beyond synchronization, we explored the dynamics of the combined system and identified regimes in which the auxiliary register becomes entangled either with the automaton or within itself. In particular, we showed that appropriately designed automata can serve as generators of multipartite entanglement, including GHZ, W, and absolutely maximally entangled (AME) states. These constructions illustrate that the structure of the underlying automaton directly influences the class of entangled states that can be produced, suggesting a deeper connection between automata theory and the classification of multipartite entanglement.

The proposed dynamics can be naturally interpreted in a quantum circuit model, where each step corresponds to a controlled unitary acting on the automaton and a single qubit of the register. This suggests that the framework could be implemented on programmable quantum processors, provided that suitable decompositions into elementary gates are available.

Our results open several directions for future research. First, it would be desirable to characterize the minimal size and complexity of automata required to generate specific classes of entangled states and to determine optimal constructions. Second, a systematic classification of unitarizable automata and their expressive power in quantum information processing remains an open problem. Third, it would be interesting to explore connections with quantum channels and dilation theory, as well as to investigate potential implementations in quantum circuits or programmable quantum processors. Finally, it would be interesting to explore the possible behavior of QDFAs in which the initial qubit registers are superpositions of classical words. Such dynamics would lead to even more complex forms of entanglement between the automaton and the qubits.

More broadly, the framework introduced here suggests that concepts from classical automata theory can be meaningfully extended to quantum settings, providing new tools for state manipulation, entanglement generation, and information encoding. We hope that this work stimulates further exploration at the interface of quantum information and theoretical computer science.


\vspace{1cm}
{\it Acknowledgements.} This research is supported by the Polish National Science Centre (NCN) under the Maestro Grant no. DEC-2019/34/A/ST2/00081.


\bibliography{main}

\end{document}